\documentclass[reprint,aps,prl]{revtex4-1}
\usepackage{graphicx}
\usepackage{bm}
\usepackage{color}
\usepackage{subfigure}
\usepackage{ulem}
\usepackage{amsmath}
\usepackage{braket}
\usepackage{hyperref}

\begin{document}
\title{Universal Features of Landau Fans of Twisted Bilayer Graphene with Large Superlattices}
\author{Tin-Lun Ho and Cheng Li}
\affiliation{Department of Physics, The Ohio State University, Columbus, OH 43210, USA} 
\date{\today}

\begin{abstract}
Current experiments on different samples of twisted bilayer graphene (TBG) have found different sets of insulating phases. Despite this diversity, many features of these insulating phases appear to be universal. They include the dispersion of Landau fans away from charge neutrality, a reduced  Landau fan degeneracy from the expected value at charge neutrality, and the further reduction of this degeneracy when crossing an insulating phase with odd number of electrons in the superlattice unit cell. We point out  that all these behaviors as well as the ferromagnetic behavior observed in  some of the insulating states suggest an underlying ``ideal" pattern, with different part of it realized in the different samples in different experiments.  We further show that such pattern can be accounted for by a Hubbard like model for the superlattice {\em augmented with  a set of  chemical potential dependent mean fields} that break the symmetry of the eight internal degrees of freedoms successively. The simultaneous importance of Mott like physics and mean field physics may be a general feature of twisted 2D electronic materials with large superlattices, not necessarily confined to graphene. 
\end{abstract}
\maketitle

Twisted bilayer graphene (TBG) at a small twist angle is full of puzzles and surprises.  Through twisting to the ``magic" angle \cite{Cao2018SC, Cao2018Ins, Lu2019} or through out-of-plane pressure \cite{Yankowitz2018}, superconductivity can be activated. 
At the same time, its normal state also displays a whole host  of baffling phenomena which seem to be related to each other \cite{Cao2018SC,Cao2018Ins, Yankowitz2018, Sharpe2019, Polshyn2019, Lu2019}.
First of all, a sequence of insulating phases are found on  both sides of charge neutrality (CN) at integer filling, i.e. integer  number of electrons per superlattice unit cell.  The sequence of insulating phases observed appears to be sample dependent
\cite{Cao2018SC,Cao2018Ins, Yankowitz2018, Sharpe2019,Polshyn2019, Lu2019}.
Secondly, the sizes of the Fermi surfaces on different sides of each insulating phases 
revealed from the Landau fans  are dramatically different. In all cases, the Landau fans  disperse in a direction away from   CN. Thirdly, the degeneracy of Landau fan near CN is  reduced from the expected value of  8  (two spins, two valleys, and two layers) \cite{Cao2016} to 4 \cite{Cao2018SC, Cao2018Ins, Yankowitz2018,Lu2019}; and continues to decrease  to 1 as one passes through the successive insulating phases to the bottom and to the top of the flat band \cite{Yankowitz2018, Lu2019}.  In all cases,  a change of degeneracy by a factor of 2 takes place whenever an insulating phase with odd number of electrons per superlattice unit cell is crossed.
The recurring pattern and the systematic reduction of degeneracy of the Landau fan suggests that similar physical mechanisms are at work as the insulating phases are crossed successively. Even though the number of insulating states varies between samples, the features mentioned above  appear to be universal. 

The insulating phases are surely caused by interactions. The question is they are consequences of  mean field physics or  strong (Mott-like)  correlations. The success of the continuum theory \cite{Bistritzer2011} in predicting  band flattening around $1.05^\circ$ shows some aspects of band theory are essential, even though it predicts a Landau fan degeneracy of 8, twice of the observed value. 
% (Recently, we have  showed  within continuum theory that the coupling between unlike valley Dirac cones in differen layers will simultaneously reduce the Landau fan degeneracy from 8 to 4 and  open an insulating gap at half filling $n/n_{s}=\pm 1/2$, where $n_s=4$ \cite{HL1}). 
On the other hand, the emergence  of  insulating phases at integer number of electrons per superlattice unit cell is most efficiently accounted for by Mott physics. As a matter of fact, if one were to generate insulating phases in  continuum theory through different types of density wave formation, one would obtain many insulating phases not necessarily at  integer filling, due to  different  types of nesting vectors. 
Hence, while some aspects of the normal state are readily explained by continuum theory, others are easily explained by the opposite Mott physics.  Neither viewpoint can provide simple  explanation for the systematic reduction of Landau fan degeneracy. 

There has been an avalanche of theoretical studies since the announcement of superconductivity in TBG.
Current theoretical studies have spanned both viewpoints. In particular, Ref.\cite{Yuan2018} and \cite{Lian2018} have addressed the issue of Landau fan degeneracy. In Ref.\cite{Yuan2018}, lattice distortion and  coulomb interaction are suggested to be the cause for  reducing  the Landau fan  degeneracy near CN  from 8 to 4; and Mott physics is invoked for the formation of the insulating phases  at carrier density  $n=\pm n_{s}/2$, where $n_s = 4$ electrons per superlattice unit cell. However, the reduction of Landau level degeneracy from 4 to 2 around the insulating phases at $|n| = n_s/2$ was not discussed. On the other hand, Ref.\cite{Lian2018} makes use of  the continuum theory. The reduction of degeneracy from 8 to 4 is attributed to Zeeman effect, and its further reduction from 4 to 2 is found at sufficiently high magnetic field.  
This picture makes no reference to  the insulating phases at $n=\pm n_{s}/2$  at zero field. So the suggested mechanism for degeneracy reduction is not tied to the physics of the insulators. It is not clear at the moment how the Landau fan degeneracy of  the  insulating phases at   $n=n_{s}/4$ and $\pm 3n_{s}/4$ \cite{Lu2019, Yankowitz2018, Sharpe2019} are explained by these pictures. 

{\em Differences between the Hubbard models of large superlattices and those of simple lattices:} If one were to derive a Hubbard model to describe the flat band physics of superlattices with large unit cells, one would expect the  hopping integral between the Wannier states of the superlattice would have included  some  interaction effects. At the same time, there will also be residential  interactions between Wannier states, such as charging energy within each superlattice unit cell (analogous to the usual Hubbard $U$), as well as effective interactions between different internal degrees of freedom (i.e. layer, valley, and spin). Since the spatial extent of the Wannier state includes hundreds of original unit cells, one expects  these effective interactions will be chemical potential dependent. Since the charging energy is electrostatic in nature (in a dielectric medium determined by many-body physics), it will retain the usual charging form $q^2/2C$ within each unit cell, where $q$ is the charge and $C$ is the capacitance of the cell. On the other hand, the residual interaction between different degrees of freedom can be mean field like.   The purpose of this paper is to show that the puzzling behavior of the Landau fans observed over the entire density regime (summarized below) can be captured by a generalized Hubbard model that  includes  a set of very simple chemical  dependent mean fields. 

{\em Summary of experimental situation:} Before proceeding, we first summarize the current  experimental findings schematically in Fig. 1. This figure shows a particle-hole symmetric pattern. It is 
identical to the result in  Ref.\cite{Yankowitz2018} except for  the filling interval $-1 /2<n/n_s < -1 /4$. 
We shall use this figure to illustrate the key findings in different experiments, as well as the working of our model.   
For TBG, the flat band near CN can be viewed as tight binding model 
on the superlattice with electrons carrying 
 8 degrees of freedom: layer index up and down $(u,d)$, sublattice (or valley) index $(K,K')$,  and spin index $(\uparrow, \downarrow)$.  The allowed number of electron ($q$) per superlattice site  ranges from 8 to 0, where  $q=8$ and $q=0$ correspond to the full band and the empty band. 
Charge neutrality  (CN) is at $q=4$.  The carrier density is $n=q-4$, and the filling is $n/n_{s}=(q/4)-1$, ($n_s=4$).  The experimental findings are: 

\vspace{0.1in}

\noindent $({\bf 1})$ {\bf Location of the insulating phases}:  All insulating phases discovered so far occur at integer number of electrons per (superlattice) site, i.e. $q=$ integer. 
In Ref.\cite{Cao2018Ins, Cao2018SC}, only  the  insulating phases at $q=0, 2, 4, 6, 8$ are observed. 
%{\color{red}\sout{In Ref.\cite{Yankowitz2018},   insulating phases at all integer  values of $q$ except for $q=3$, or $n=-n_{s}/4$ are found. } 
In the absence of magnetic field, insulating phases at all integer values of $q$ except for $q=3$ (or $n=-n_{s}/4$) are found \cite{Yankowitz2018, Lu2019}. The $q=3$ insulating state is reported in \cite{Lu2019} when applying magnetic field.
 Another recent experiment (Ref.\cite{Sharpe2019})  has only found insulating state above charge neutrality, at $q=4, 6, 7$, besides the band insulator at $q = 0$ and  8.  In Fig.1, we show the case where insulating phases occur at all integer $q$. 

\noindent $({\bf 2})$  {\bf Landau fan degeneracy ($G$) around  CN}: 
Current experiments show that the Landau fans emerge symmetrically on both sides of CN.  In
Ref.\cite{Cao2018Ins, Cao2018SC, Yankowitz2018, Lu2019}, these Landau fans have degeneracy $G=4$ instead of 8 predicted by band theory \cite{Bistritzer2011} (See Fig. 1). In Ref. \cite{Sharpe2019}, this degeneracy is even reduced to 2.  Despite this difference, the reduction of degeneracy from the expected value of 8 occurs in all current experiments. 

\noindent $({\bf 3})$ {\bf Landau fans  away from  CN}: 
As filling $n$ increases above  CN, the Landau fan expands until it reaches an insulating phase. On the other side of the insulator, a new Landau fan emerges. 
The Landau fans on both sides of an insulating phase indicate a large Fermi surface at fillings below it  and a small Fermi surface above it.
 An opposite  situation is  found for fillings below CN\cite{Cao2018Ins, Cao2018SC, Yankowitz2018, Sharpe2019, Lu2019} (See Fig.1).  That all Landau fans disperse away from CN, and the dramatic changes in Fermi surface area after crossing an insulating phase appear to be universal. 

\noindent $({\bf 4})$ {\bf Degeneracy patterns away from CN} :  In Ref.\cite{Yankowitz2018}, the 
 degeneracy of the Landau fans  $(G)$ varies as   $ 4, 2, 2 ,1$  and $4,2,1$ as one sweeps  through the metallic phases above and below  CN. Moreover, the reduction of $G$ appears at insulating states with odd $q$'s. Fig.1 shows the particle-hole symmetric version of the finding of  Ref.\cite{Yankowitz2018}, with the pattern of $G$  (4,2,2,1)  on both sides of CN. 
Experiments that observed fewer insulating states have fewer Landau fans. That   $G$ is reduced by 2 across insulating phases with odd $q$, and remain constant across those with even $q$ also appears to be universal \cite{Cao2018Ins, Cao2018SC, Yankowitz2018, Lu2019, Sharpe2019}.

\noindent $({\bf 5})$ {\bf Magnetic field effects }: At sufficiently large magnetic fields, the degeneracy of  Landau fans is found to reduce  in the  interval $(2<q<4)$ % ( $-n_{s}/2< n < 0$ ) 
in Ref.\cite{Yankowitz2018}.  

\noindent $({\bf 6})$ {\bf Ferromagnetism}: Anomalous  Hall effect is found at $q=7$ state in Ref.\cite{Sharpe2019}. It was taken as  evidence of ferromagnetism. 

\noindent $({\bf 7})$ {\bf Landau fans associated with non-integer $q$}: In Ref. \cite{Yankowitz2018}, a Landau fan is found to converge to  $n= -2n_{s}/3$, corresponding to non-integer number of electrons in each superlattice unit cell. 

We shall  present below a simple model that can account for $({\bf 1})$ to $({\bf 6})$. In a separate paper \cite{HL1}, we shall show that in the weak coupling, density wave states triggered by nesting effects can  lead to 
Landau fans at  non-integer $q$ as in  $({\bf 7})$.

\begin{figure}[htbp]
\includegraphics[width =3.2in]{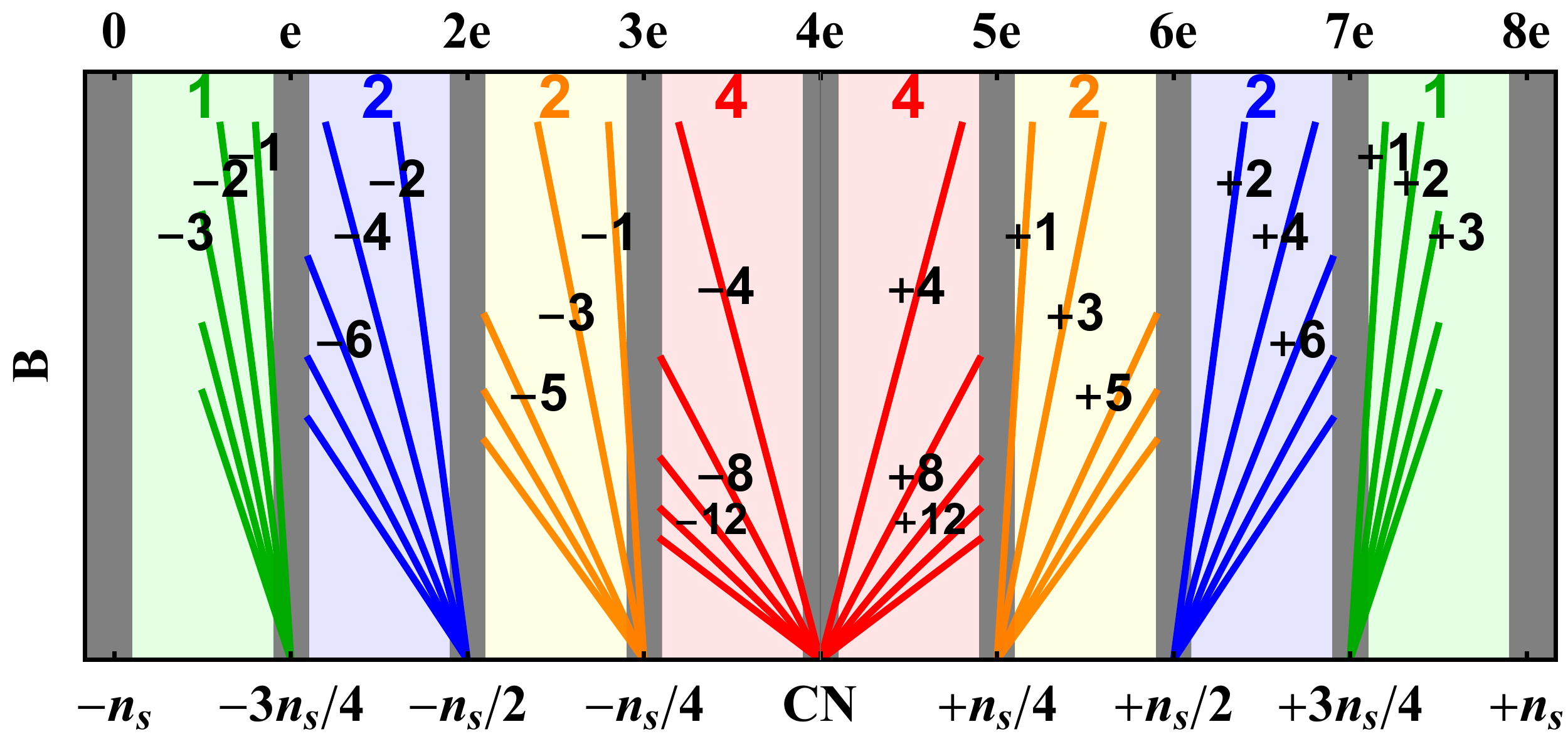}
\caption{A particle-hole symmetric idealization of the experimental result  in Ref.\cite{Yankowitz2018}. The lower axis is  carrier density $n$, and the upper axis is the number of changes per superlattice site, $qe$. The regions where the Landau fans appear are metallic regions. The large numeral denotes the degeneracy $G$ of the Landau fan. 
Although different experiments  show different  subset of this pattern, the features $({\bf 2})$ to $({\bf 4})$ discussed in text appear to be universal. }
\label{fig:LF}
\end{figure}

\begin{figure}[htbp]
\center
\includegraphics[width = 3.2in]{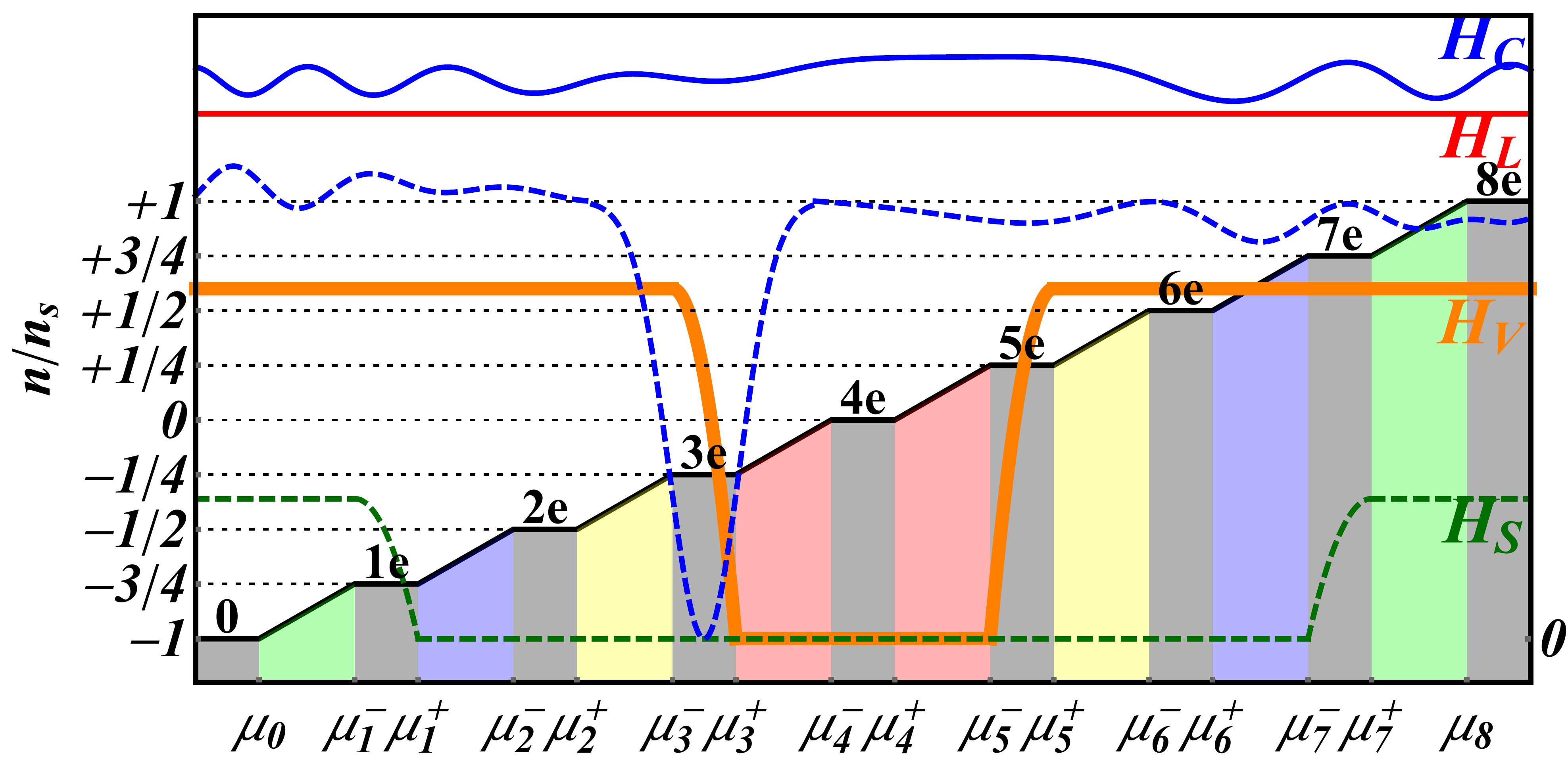}
\caption{ The chemical potential dependence of various interactions. The charging energy $H_{c}$ is represented by a solid blue  line. Like $H_{c}$, the mean field $H_{L}$ causing layer polarization is turned on for the entire range of $\mu$. The mean field $H_{V}$ causing  valley polarization is turned on at $\mu^{+}_{3}$ and $\mu^{-}_{5}$, and 
the mean field $H_{S}$  causing  spin polarization is turned on at $\mu^{+}_{1}$ and $\mu^{-}_{7}$. The flat portion of these mean fields is for easy visualization. In general they can  vary with $\mu$. In the case $H_{c}=q^2/2C$, if the capacitance becomes very big in the interval $\mu^{-}_{3}<\mu < \mu^{+}_{3}$, $H_{c}$ 
(represented by the dashed curve) 
%(blue dashed curve) 
will vanish around that interval and the $q=3$ Mott state will be absent. }
\end{figure}

{\em Our model:}  With our earlier discussions, we  consider a Hubbard model consists of the usual hopping $\hat{T}$, a charging energy $H_{c} = q^2/2C= (e^2/2C)n^2 \equiv Un^2$, and  a set  of mean fields that  set in at different chemical potentials $\mu$,  favoring certain  structures of internal degrees of freedom.  
We shall first consider  interactions that lead to a particle-hole symmetric Landau fan pattern shown in Fig. 1, and then show how to obtain the features $({\bf 1})$-$({\bf 6})$. 
We shall consider the  $\mu$-dependent  mean field  $H_L$, $H_V$, and $H_S$  that favor specific configurations in layer, valley, and spin spaces separately.  (We {\em stress}  that such  separation is chosen to simplify  the discussions. Our results  remain unchanged even when the mean fields couple different types of degrees of freedom.)
We shall denote the states favored and disfavored  in layer space ${\cal L}\equiv (u,d)$ as ${\cal U}$ and ${\cal D}$, those in 
valley space ${\cal V}\equiv (K,K')$ as ${\cal K}$ and $ {\cal K}'$, and those in spin ${\cal S}\equiv (\uparrow, \downarrow)$ spaces as $\alpha$ and $ \beta$. The pairs $({\cal U}, {\cal D})$, $({\cal K}, {\cal K}')$, and $(\alpha, \beta)$  are rotations of $(u,d)$, $(K, K')$, and 
$(\uparrow, \downarrow)$ respectively. These interactions are turned on as follows, and are shown in Fig. 2 : 

\noindent  $( {\bf A})$ The charging energy $H_{c}$ is present for all   $\mu$.  The capacitance $C$ can be $\mu$-dependent. $H_{c}$ will  produce Mott insulators at integer number per cell  $q=1,2, \ldots, 7$.
It is invariant with respect to rotations  in layer ${\cal L}$, valley ${\cal V}$ and spin ${\cal S}$ spaces, and is therefore 8-fold degenerate.
The upper and lower bound of the chemical potential of the  $q$-particle Mott insulator will be denoted as $\mu_{q}^{-}$ and $\mu_{q}^{+}$.

\noindent $( {\bf B})$ A symmetry breaking field $H_{L}$ in the layer space ${\cal L}$ is also present for all   $\mu$ favoring a layer state ${\cal U}$.  $H_{L}$  is independent of  valley 
and   spin  indices  and is 4-fold degenerate. 
(A similar situation occurs in the weak coupling case. The 
coupling of unlike valley Dirac cones  from different layers can produce a layer polarization that reduces the Landau fan degeneracy from 8 to 4, while opening a gap  at $n=\pm n_{s}/2$ simultaneously.\cite{HL1}). 

\noindent $( {\bf C})$ As $\mu$ increases above CN, new mean fields are set in  at the Mott phases with odd particle numbers $q$ per site. At the onset of the $q=5$ Mott state, $\mu^{-}_{5}$,  a mean field  $H_{V}$ favoring the valley state ${\cal K}$ is turned on as shown in Fig. 2, gradually establishing an ordering 
in state ${\cal K}$ within the insulating phase ($\mu^{-}_{5}<\mu <\mu^{+}_{5}$).  
Like $H_{c}$, $H_{V}$ does not affect the spin degrees of freedom and has a degeneracy of 2. 

\noindent $( {\bf D})$ As $\mu$ increases further to $\mu^{-}_{7}$, the onset of the $n=7$ Mott state, 
a mean field  $H_{S}$ favoring the  spin state $\alpha$ is turned on as shown in Fig. 2, 
gradually  polarizing the electrons to spin $\alpha$ within the interval ($\mu^{-}_{7}<\mu <\mu^{+}_{7}$).  (See Fig 2.)

\noindent $({\bf E})$ Similar symmetry breaking fields are switched on in a particle-hole symmetric manner for $\mu$  below CN,  i.e. $-H_{V}$ is switched on at $\mu_{3}^{+}$  favoring the state ${\cal K'}$ within the range $\mu_{3}^{-}<\mu < \mu_{3}^{+}$, and $-H_{S}$ is  switched on at $\mu_{1}^{+}$  favoring the state $\beta$ within the range $\mu_{1}^{-}<\mu < \mu_{1}^{+}$ as shown in Fig. 2. 

Although we have chosen to first break the valley symmetry and then the spin symmetry as $\mu$ moves away from   CN,  the order can be changed in either  or both sides of  CN without affecting our  conclusions.  We now show that the  interactions $({\bf A})-({\bf E})$  lead to the pattern in Fig. 1 : 

\begin{figure}[htbp]
\centering
\includegraphics[width=3.2in]{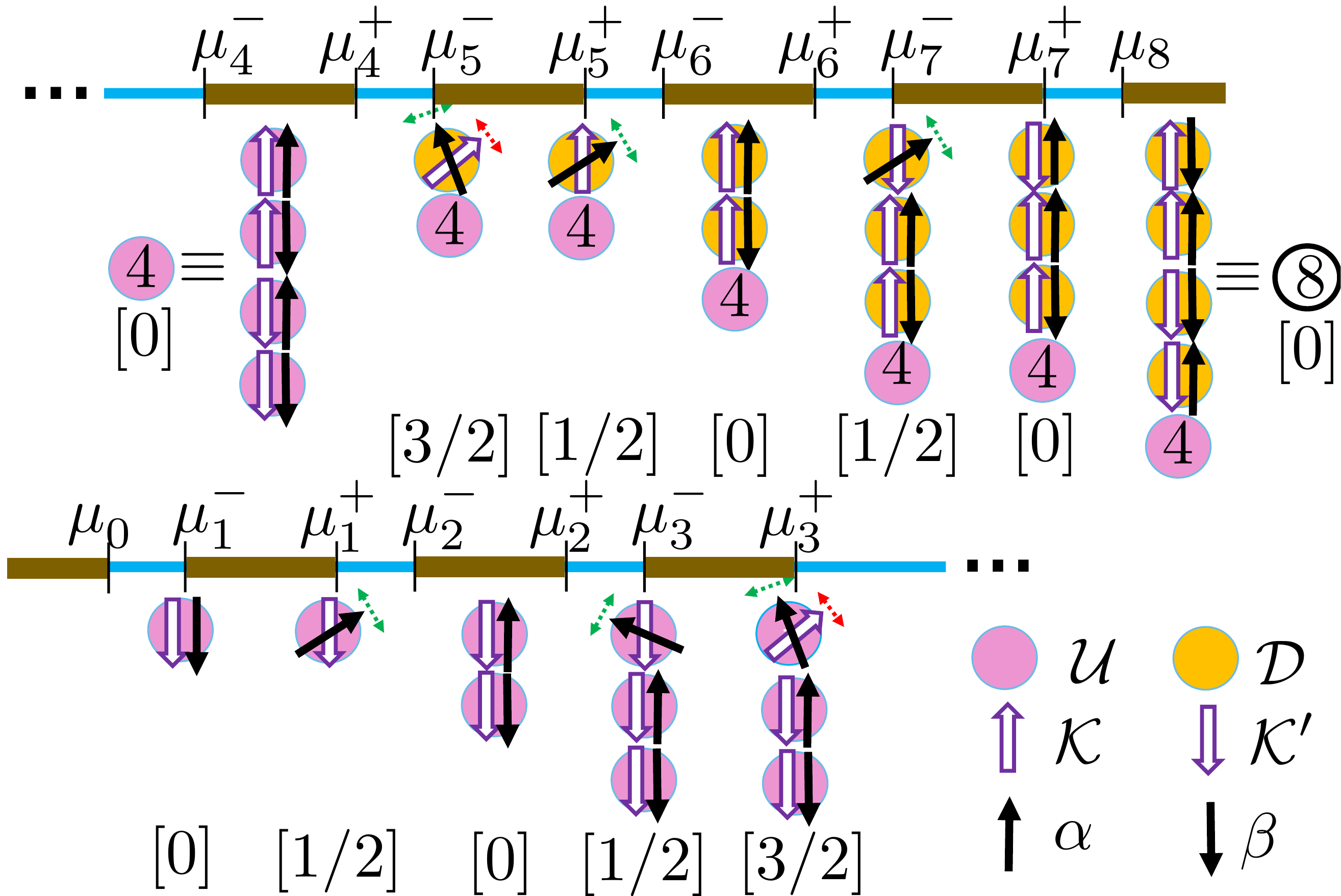} 
\caption{The sequence of Mott states generated by our model: The stack of circles is to represent that Fock state (or number state) for fermions at each insulator at each site $|q\rangle_{\bf R}$ on the superlattice as discussed in text. 
The states ${\cal U}$ and ${\cal D}$ are represented by different colors. The 
white and black arrow represent  valley and spin degrees of freedom.  A tilted back arrow associated with a perpendicular dashed two-head arrow outside means that the spin can be in any state. When spin is polarized by the mean field, the arrow points either up or down.  The same applies to the valley degrees of freedom (white arrow). For $q\leq 4$, all sites are occupied by the ${\cal U}$ state. The Mott state $q \geq 5$ will have  additional ${\cal D}$ particles. At $\mu^{-}_{5}$, the added particle has full valley and spin degeneracy, and behaves like a spin-3/2 fermion, denoted as [3/2]. As $\mu$ increases from $\mu^{-}_{5}$ to $\mu^{+}_{5}$, the valley polarization $H_{V}$ is turned on, freezing the valley degrees of freedom and turing it into a  spin-1/2 particle.   Similar reduction of degenerate degrees of freedom also occurs in the other odd $q$ states.  The Mott state $|4\rangle$ is a valley and spin singlet. The Mott state $|2\rangle$ and $|6\rangle$ are valley ferromgnet and spin singlets. The Mott states  $|1, \mu^{-}_{1}\rangle$ and $|7, \mu^{+}_{7}\rangle$ are valley and spin ferromagnets.  The Mott state at $\mu^{+}_{1}, \mu^{-}_{3},   \mu^{+}_{5}, \mu^{-}_{7}$ are the ground state of a spin-1/2 antiferromagnet Heisenberg hamiltonian.   The Mott state at $\mu^{+}_{3}$ and $\mu^{-}_{5}$ are the ground state of a spin-3/2 antiferromagnet Heisenberg hamiltonian. The particle and hole excitations of these states will give rise to a Landau fan degeneracy $4,2,2, 1$ away from CN as explained in $({\bf I})$ to  $({\bf V})$  in text.}
\label{fig:fill}
\end{figure}

\noindent  ({\bf I})   We shall denote the Mott state  with $q$ electrons per site as $|q\rangle$. Both ({\bf A}) and  ({\bf B}) imply  that  the Mott states with $q\leq 4$ are make up of electrons in the ${\cal U}$ state. 
The creation operator for the ${\cal U}$ state and the orthogonal   ${\cal D}$ state  are denoted as  
  $\hat{\cal U}^{\dagger}_{\ell}$ and  $\hat{\cal D}^{\dagger}_{\ell}$, where $\ell$  labels the four states $(K \uparrow)$, $(K' \downarrow)$, $(K \uparrow)$, $(K' \downarrow)$. The Mott state at CN is  $|4\rangle= \prod_{\bf R} |4\rangle_{\bf R}$, where $|4\rangle_{\bf R} = 
\prod_{\ell = 1}^{4} \hat{\cal U}^{\dagger}_{\ell}({\bf R})|0\rangle$ is a valley and spin singlet. See Fig. 3. 
Its particle and hole excitations are the momentum ${\bf k}$ states ${\cal D}^{\dagger}_{\ell}({\bf k})|4\rangle$ and 
 ${\cal U}^{}_{\ell}({\bf k})|4\rangle$.  
 Since $H_{L}$ is independent of $\ell$, these excitations are 4-fold degenerate and behave like a spin-3/2 particles. 
As $\mu$ moves away from CN,  these ``spin=3/2"  particles gather into a growing Fermi surface ,  leading to an expanding Landau fan with $G=4$ on both sides of CN as shown in Fig.1  in the  interval  $-n_{s}/4<n< n_{s}/4$. 
It is useful to think of the Mott state with $q$ as $q=4+n$ state, where $n = -4, \ldots, +4$, with the state $|4\rangle$ being an inert background, since it is a valley and spin singlet.

\noindent  ({\bf II}) 
As $\mu$ approaches  $\mu^{-}_{5}$ from below, one reaches the onset of the $q=5=4+1$ (or $n=n_{s}/4$ Mott state). The {\em large} Fermi sea of  the ``spin-3/2" fermions turns into a Mott insulator of spin-3/2 particles. (See Fig.3)
Each site is in the state  
$|5, \ell \rangle_{\bf R} = \hat{D}^{\dagger}_{\ell}({\bf R})|4\rangle_{\bf R}$. 
In the absence of  symmetry breaking fields in the space ${\cal V}\otimes{\cal S}$, super-exchange will generate  a Heisenberg like Hamiltonian for this system. The Mott state  is then 
$|5, \mu_{5}^{-}\rangle =  \hat{O}^{\dagger}(\mu^{-}_{5})|4\rangle$, where 
$\hat{O}^{\dagger}(\mu^{-}_{5}) = \sum_{[\ell]} \Psi_{\ell_1, \ell_{2}, ..\ell_{N} }\prod_{{\bf R}_{i}} \hat{\cal D}^{\dagger}_{\ell_i} ({\bf R}_{i})$, and $\Psi$ is the ground state of the relevant spin-3/2 Heisenberg Hamiltonian, which we shall take as a singlet in  ${\cal V}\otimes{\cal S}$. 
However, as $\mu$ increases further, the  mean field $H_{V}$ discussed in $({\bf C})$ grows. 
 It favors the valley state ${\cal K}$, and hence  reduces the degeneracy $\ell$  to $({\cal K}, \sigma)$, $\sigma = \uparrow, \downarrow$. This process changes 
 the spin-3/2 fermions into spin-1/2 fermions as $\mu$ sweeps  through the insulator from below, $ \mu^{-}_{5}<\mu< \mu^{+}_{5}$. 
The  ground state near $\mu^{+}_{5}$ is  
$|5, \mu_{5}^{+}\rangle =  \hat{O}^{\dagger}(\mu^{+}_{5})|4\rangle$, where 
$\hat{O}^{\dagger}(\mu^{+}_{5}) = \sum_{[\sigma]} \Phi_{\sigma_1, \sigma_{2}, ..\sigma_{N} }\prod_{{\bf R}_{i}} \hat{\cal D}^{\dagger}_{{\cal K} \sigma_i} ({\bf R}_{i})$, and $\Phi$ is the ground state of the  spin-1/2 Heisenberg model near $\mu^{+}_{5}$, which is a spin singlet.  See Fig.2 and 3. 

The electron excitation at $\mu_{5}^{+}$ is  $\hat{\cal D}^{\dagger}_{{\cal K}, \sigma}({\bf k})|5, \mu_5^+\rangle$, which is two fold degenerate and is a  spin-1/2 fermion. As $\mu$ increases above  $\mu^{+}_{5}$, these excitations form a growing Fermi surfaces starting from zero size at $\mu^{+}_{5}$,  leading to  a growing Landau fan dispersing towards the top of the band with  $G=2$. 
In contrast, the Landau fan for $\mu$ just below $\mu_{5}^{-}$ is due to the electron excitations originated from the state $|q=4\rangle$, which has a large Fermi surface at  $\mu_{5}^{-}$.  So there is a great asymmetry of the size of Fermi surface 
on different sides of $|q=5\rangle$
as shown in Fig.1.

\noindent ({\bf III})  As $\mu$ reaches $\mu^{-}_{6}$,  the onset of the $q=6=4+2$ Mott state  ($n=n_s /2$),  
the spin-1/2 excitations  emerging from $|5\rangle$ are localized, turning into  the Mott state $|6\rangle = \prod_{\bf R} 
\hat{\cal D}^{\dagger}_{ {\cal K} \uparrow} ({\bf R})\hat{\cal D}^{\dagger}_{ {\cal K} \downarrow}({\bf R})|4\rangle$ (See Fig.3). Since this  state is a valley ferromagnet and a spin singlet, its  internal structure of this state is unchanged over the range
$\mu^{-}_{6}<\mu < \mu^{+}_{6}$.  

The electron excitation of this state is ${\cal D}^{\dagger}_{{\cal K}' \sigma}({\bf k}) |2\rangle$, which is again a spin-1/2 fermion. Hence, the degeneracy of the Landau fans remains $G=2$ 
on both sides of the  (even) $q=6$  insulator, unlike the  (odd) $q=5$. 
Yet in both  cases, the size of Fermi surface jumps from large value to  0 as $\mu$ passes through an  insulating gap from below as shown in See Fig.1. 

\noindent  ({\bf IV})  As $\mu$ increases to $\mu^{-}_{7}$,  the onset of the $q=7$ Mott state $|7\rangle$ (or $n= (3/4)n_s$), 
 each site is a Fock state $|7, \sigma \rangle_{\bf R}= {\cal D}^{\dagger}_{K' \sigma}|6\rangle_{\bf R}$, which is a spin-1/2 particle. (See Fig.3).  Again, super-exchange will generate an antiferromagnet Heisenberg Hamiltonian for this system. 
 The Mott state at  $\mu^{-}_{7}$ is then 
$|7, \mu_{7}^{-}\rangle = \hat{O}^{\dagger}|6\rangle$, $\hat{O}^{\dagger} = 
\sum_{[\sigma]} \Phi_{\sigma_1, \sigma_{2}, ..\sigma_{N} }\prod_{{\bf R}_{i}} \hat{\cal D}^{\dagger}_{{\cal K}' \sigma_i} ({\bf R}_{i})$, where $\Phi$ is the antiferromagnet ground state. 
However, as mean field $H_{S}$ is turned  on with increasing  $\mu$ (see ${\bf D}$),  the ground state turns into a ferromagnet with spin  state $\alpha$, and 
$|7, \mu_{7}^{+}\rangle = \prod_{\bf R}{\cal D}^{\dagger}_{{\cal K}', \alpha}({\bf R})|6\rangle$.  The particle excitations are obtained by adding fermions in the orthogonal spin states,  ${\cal D}^{\dagger}_{{\cal K}', \beta}({\bf k})|7, \mu_7^+\rangle$. These excitations are spin, valley, and layer polarized.  They then lead to a non-degenerate Landau fan growing out from $\mu_{7}^{+}$ with $G=1$, as shown in Fig.1.

\noindent ({\bf V})  The situation below  CN is the particle-hole mirror of $({\bf I})$ to $({\bf IV})$, with the creation of the ${\cal D}$ particle  replaced by  the removal of  the ${\cal U}$ particles.  Reasoning similar to  $({\bf I})-({\bf IV})$ then lead to 
the insulating structure shown in Fig. 3, which leads to 
the degeneracy pattern $(4,2,2,1)$ on both sides of CN as shown in Fig 1.

{\em Connection to  experiments:}  We now relate the results $({\bf I})$ to $({\bf V})$ to experimental results $({\bf 1})$ to $({\bf 6})$:  

\noindent $({\bf 1'})$ Since the charging energy is sample dependent,   the absence of Mott state at $q=3$ (or $n= -n_{s}/4$) in current experiments (discussed  in  $({\bf 1})$) can occur if  the  capacitance $C$ is very  large in the range of $\mu$  for  the $q=3$ state. It remains to be seen whether this insulating state remains absent in future experiments. 
The missing insulating phase at $n=\pm n_s/4$ and $n = \pm 3n_s/4$  in Ref. \cite{Cao2018Ins, Cao2018SC} could also be explained by the same reason.
 The absence of insulating  state at densities below  CN  reported in Ref.\cite{Sharpe2019} will then correspond  to a very large $C$   for  all  $\mu$ below  CN.  

\noindent $({\bf 2'})$The observations in $({\bf 2})$-$({\bf 4})$: Our model will give rise to the observation in Ref.\cite{Yankowitz2018} provided the insulating phase at $q=3$ is eliminated (see $({\bf 1'})$). The lower degeneracy of 2 around CN observed in Ref.\cite{Sharpe2019}  can be obtained from our model  if the valley polarization $H_{V}$ that favors a particular valley state ${\cal K}$ is present  for all $\mu$, just like the layer polarization $H_{L}$.  This will freeze both layer and valley degrees of freedom, leaving spin is the only degeneracy. Our model also shows that all the universal features  in $({\bf 2})$-$({\bf 4})$. 
 
\noindent $({\bf 3'})$ 
In our model, the $q=7$ (or $n= 3n_s /4$) Mott state is a spin ferromagnet. Even though our model need not be the the reason for the  observed ferromagnetic state Ref.\cite{Sharpe2019}, (See ({\bf 6})),  it shows a pathway to ferrmagnetism in the insulating phases. Should  spin symmetry be broken before the valley symmetry as one moves away from CN, ferromagnetism will be found in more than one insulating states.

\noindent $({\bf 4}')$ {\em Zeeman Effect:} 
The wavefunctions of the degenerate states in a Landau fan generally involve mixing of layer, valley, and spin degrees of freedom. A magnetic field will deform these wavefunctions, but need not split the degeneracy if the energy of symmetry  mixing still dominates over the Zeeman energy. The critical magnetic field at which the splitting of Landau fan occurs is therefore a measure of the the strength of symmetry mixing.

 {\em Final remarks:} While we have made the point that many puzzling features of twisted bilayer graphene can be explained 
by  a simple model that  includes both strong correlation and mean field physics, the specific features of graphene invoked are only related to its the degrees of freedom of the superlattice, which is quite minimal.  Our formulation, as it stands, is not restricted to graphene. It will be interesting to test  these results in other twisted 2D materials. 
 
 Acknowledgments: The work is supported by MURI Grant FP054294-D, the NASA Grant on Fundamental Physics 1541824, 
 %NSFC grant (No. 11674192), 
 and the OSU MRSEC Seed Grant.

\bibliographystyle{apsrev4-1.bst}
\bibliography{Ref}

\begin{thebibliography}{11}%
\makeatletter
\providecommand \@ifxundefined [1]{%
 \@ifx{#1\undefined}
}%
\providecommand \@ifnum [1]{%
 \ifnum #1\expandafter \@firstoftwo
 \else \expandafter \@secondoftwo
 \fi
}%
\providecommand \@ifx [1]{%
 \ifx #1\expandafter \@firstoftwo
 \else \expandafter \@secondoftwo
 \fi
}%
\providecommand \natexlab [1]{#1}%
\providecommand \enquote  [1]{``#1''}%
\providecommand \bibnamefont  [1]{#1}%
\providecommand \bibfnamefont [1]{#1}%
\providecommand \citenamefont [1]{#1}%
\providecommand \href@noop [0]{\@secondoftwo}%
\providecommand \href [0]{\begingroup \@sanitize@url \@href}%
\providecommand \@href[1]{\@@startlink{#1}\@@href}%
\providecommand \@@href[1]{\endgroup#1\@@endlink}%
\providecommand \@sanitize@url [0]{\catcode `\\12\catcode `\$12\catcode
  `\&12\catcode `\#12\catcode `\^12\catcode `\_12\catcode `\%12\relax}%
\providecommand \@@startlink[1]{}%
\providecommand \@@endlink[0]{}%
\providecommand \url  [0]{\begingroup\@sanitize@url \@url }%
\providecommand \@url [1]{\endgroup\@href {#1}{\urlprefix }}%
\providecommand \urlprefix  [0]{URL }%
\providecommand \Eprint [0]{\href }%
\providecommand \doibase [0]{http://dx.doi.org/}%
\providecommand \selectlanguage [0]{\@gobble}%
\providecommand \bibinfo  [0]{\@secondoftwo}%
\providecommand \bibfield  [0]{\@secondoftwo}%
\providecommand \translation [1]{[#1]}%
\providecommand \BibitemOpen [0]{}%
\providecommand \bibitemStop [0]{}%
\providecommand \bibitemNoStop [0]{.\EOS\space}%
\providecommand \EOS [0]{\spacefactor3000\relax}%
\providecommand \BibitemShut  [1]{\csname bibitem#1\endcsname}%
\let\auto@bib@innerbib\@empty
%</preamble>
\bibitem [{\citenamefont {Cao}\ \emph {et~al.}(2018{\natexlab{a}})\citenamefont
  {Cao}, \citenamefont {Fatemi}, \citenamefont {Fang}, \citenamefont
  {Watanabe}, \citenamefont {Taniguchi}, \citenamefont {Kaxiras},\ and\
  \citenamefont {Jarillo-Herrero}}]{Cao2018SC}%
  \BibitemOpen
  \bibfield  {author} {\bibinfo {author} {\bibfnamefont {Y.}~\bibnamefont
  {Cao}}, \bibinfo {author} {\bibfnamefont {V.}~\bibnamefont {Fatemi}},
  \bibinfo {author} {\bibfnamefont {S.}~\bibnamefont {Fang}}, \bibinfo {author}
  {\bibfnamefont {K.}~\bibnamefont {Watanabe}}, \bibinfo {author}
  {\bibfnamefont {T.}~\bibnamefont {Taniguchi}}, \bibinfo {author}
  {\bibfnamefont {E.}~\bibnamefont {Kaxiras}}, \ and\ \bibinfo {author}
  {\bibfnamefont {P.}~\bibnamefont {Jarillo-Herrero}},\ }\href@noop {}
  {\bibfield  {journal} {\bibinfo  {journal} {Nature}\ }\textbf {\bibinfo
  {volume} {556}},\ \bibinfo {pages} {43} (\bibinfo {year}
  {2018}{\natexlab{a}})}\BibitemShut {NoStop}%
\bibitem [{\citenamefont {Cao}\ \emph {et~al.}(2018{\natexlab{b}})\citenamefont
  {Cao}, \citenamefont {Fatemi}, \citenamefont {Demir}, \citenamefont {Fang},
  \citenamefont {Tomarken}, \citenamefont {Luo}, \citenamefont
  {Sanchez-Yamagishi}, \citenamefont {Watanabe}, \citenamefont {Taniguchi},
  \citenamefont {Kaxiras}, \citenamefont {Ashoori},\ and\ \citenamefont
  {Jarillo-Herrero}}]{Cao2018Ins}%
  \BibitemOpen
  \bibfield  {author} {\bibinfo {author} {\bibfnamefont {Y.}~\bibnamefont
  {Cao}}, \bibinfo {author} {\bibfnamefont {V.}~\bibnamefont {Fatemi}},
  \bibinfo {author} {\bibfnamefont {A.}~\bibnamefont {Demir}}, \bibinfo
  {author} {\bibfnamefont {S.}~\bibnamefont {Fang}}, \bibinfo {author}
  {\bibfnamefont {S.~L.}\ \bibnamefont {Tomarken}}, \bibinfo {author}
  {\bibfnamefont {J.~Y.}\ \bibnamefont {Luo}}, \bibinfo {author} {\bibfnamefont
  {J.~D.}\ \bibnamefont {Sanchez-Yamagishi}}, \bibinfo {author} {\bibfnamefont
  {K.}~\bibnamefont {Watanabe}}, \bibinfo {author} {\bibfnamefont
  {T.}~\bibnamefont {Taniguchi}}, \bibinfo {author} {\bibfnamefont
  {E.}~\bibnamefont {Kaxiras}}, \bibinfo {author} {\bibfnamefont {R.~C.}\
  \bibnamefont {Ashoori}}, \ and\ \bibinfo {author} {\bibfnamefont
  {P.}~\bibnamefont {Jarillo-Herrero}},\ }\href@noop {} {\bibfield  {journal}
  {\bibinfo  {journal} {Nature}\ }\textbf {\bibinfo {volume} {556}},\ \bibinfo
  {pages} {80} (\bibinfo {year} {2018}{\natexlab{b}})}\BibitemShut {NoStop}%
\bibitem [{\citenamefont {Lu}\ \emph {et~al.}(2019)\citenamefont {Lu},
  \citenamefont {Stepanov}, \citenamefont {Yang}, \citenamefont {Xie},
  \citenamefont {Aamir}, \citenamefont {Das}, \citenamefont {Urgell},
  \citenamefont {Watanabe}, \citenamefont {Taniguchi}, \citenamefont {Zhang},
  \citenamefont {Bachtold}, \citenamefont {MacDonald},\ and\ \citenamefont
  {Efetov}}]{Lu2019}%
  \BibitemOpen
  \bibfield  {author} {\bibinfo {author} {\bibfnamefont {X.}~\bibnamefont
  {Lu}}, \bibinfo {author} {\bibfnamefont {P.}~\bibnamefont {Stepanov}},
  \bibinfo {author} {\bibfnamefont {W.}~\bibnamefont {Yang}}, \bibinfo {author}
  {\bibfnamefont {M.}~\bibnamefont {Xie}}, \bibinfo {author} {\bibfnamefont
  {M.~A.}\ \bibnamefont {Aamir}}, \bibinfo {author} {\bibfnamefont
  {I.}~\bibnamefont {Das}}, \bibinfo {author} {\bibfnamefont {C.}~\bibnamefont
  {Urgell}}, \bibinfo {author} {\bibfnamefont {K.}~\bibnamefont {Watanabe}},
  \bibinfo {author} {\bibfnamefont {T.}~\bibnamefont {Taniguchi}}, \bibinfo
  {author} {\bibfnamefont {G.}~\bibnamefont {Zhang}}, \bibinfo {author}
  {\bibfnamefont {A.}~\bibnamefont {Bachtold}}, \bibinfo {author}
  {\bibfnamefont {A.~H.}\ \bibnamefont {MacDonald}}, \ and\ \bibinfo {author}
  {\bibfnamefont {D.~K.}\ \bibnamefont {Efetov}},\ }\href@noop {} {\bibfield
  {journal} {\bibinfo  {journal} {arXiv:1903.06513}\ } (\bibinfo {year}
  {2019})}\BibitemShut {NoStop}%
\bibitem [{\citenamefont {Yankowitz}\ \emph {et~al.}(2019)\citenamefont
  {Yankowitz}, \citenamefont {Chen}, \citenamefont {Polshyn}, \citenamefont
  {Zhang}, \citenamefont {Watanabe}, \citenamefont {Taniguchi}, \citenamefont
  {Graf}, \citenamefont {Young},\ and\ \citenamefont {Dean}}]{Yankowitz2018}%
  \BibitemOpen
  \bibfield  {author} {\bibinfo {author} {\bibfnamefont {M.}~\bibnamefont
  {Yankowitz}}, \bibinfo {author} {\bibfnamefont {S.}~\bibnamefont {Chen}},
  \bibinfo {author} {\bibfnamefont {H.}~\bibnamefont {Polshyn}}, \bibinfo
  {author} {\bibfnamefont {Y.}~\bibnamefont {Zhang}}, \bibinfo {author}
  {\bibfnamefont {K.}~\bibnamefont {Watanabe}}, \bibinfo {author}
  {\bibfnamefont {T.}~\bibnamefont {Taniguchi}}, \bibinfo {author}
  {\bibfnamefont {D.}~\bibnamefont {Graf}}, \bibinfo {author} {\bibfnamefont
  {A.~F.}\ \bibnamefont {Young}}, \ and\ \bibinfo {author} {\bibfnamefont
  {C.~R.}\ \bibnamefont {Dean}},\ }\href@noop {} {\bibfield  {journal}
  {\bibinfo  {journal} {Science}\ }\textbf {\bibinfo {volume} {363}},\ \bibinfo
  {pages} {1059} (\bibinfo {year} {2019})}\BibitemShut {NoStop}%
\bibitem [{\citenamefont {Sharpe}\ \emph {et~al.}(2019)\citenamefont {Sharpe},
  \citenamefont {Fox}, \citenamefont {Barnard}, \citenamefont {Finney},
  \citenamefont {Watanabe}, \citenamefont {Taniguchi}, \citenamefont
  {Kastner},\ and\ \citenamefont {Goldhaber-Gordon}}]{Sharpe2019}%
  \BibitemOpen
  \bibfield  {author} {\bibinfo {author} {\bibfnamefont {A.~L.}\ \bibnamefont
  {Sharpe}}, \bibinfo {author} {\bibfnamefont {E.~J.}\ \bibnamefont {Fox}},
  \bibinfo {author} {\bibfnamefont {A.~W.}\ \bibnamefont {Barnard}}, \bibinfo
  {author} {\bibfnamefont {J.}~\bibnamefont {Finney}}, \bibinfo {author}
  {\bibfnamefont {K.}~\bibnamefont {Watanabe}}, \bibinfo {author}
  {\bibfnamefont {T.}~\bibnamefont {Taniguchi}}, \bibinfo {author}
  {\bibfnamefont {M.~A.}\ \bibnamefont {Kastner}}, \ and\ \bibinfo {author}
  {\bibfnamefont {D.}~\bibnamefont {Goldhaber-Gordon}},\ }\href@noop {}
  {\bibfield  {journal} {\bibinfo  {journal} {arXiv:1901.03520}\ } (\bibinfo
  {year} {2019})}\BibitemShut {NoStop}%
\bibitem [{\citenamefont {Polshyn}\ \emph {et~al.}(2019)\citenamefont
  {Polshyn}, \citenamefont {Yankowitz}, \citenamefont {Chen}, \citenamefont
  {Zhang}, \citenamefont {Watanabe}, \citenamefont {Taniguchi}, \citenamefont
  {Dean},\ and\ \citenamefont {Young}}]{Polshyn2019}%
  \BibitemOpen
  \bibfield  {author} {\bibinfo {author} {\bibfnamefont {H.}~\bibnamefont
  {Polshyn}}, \bibinfo {author} {\bibfnamefont {M.}~\bibnamefont {Yankowitz}},
  \bibinfo {author} {\bibfnamefont {S.}~\bibnamefont {Chen}}, \bibinfo {author}
  {\bibfnamefont {Y.}~\bibnamefont {Zhang}}, \bibinfo {author} {\bibfnamefont
  {K.}~\bibnamefont {Watanabe}}, \bibinfo {author} {\bibfnamefont
  {T.}~\bibnamefont {Taniguchi}}, \bibinfo {author} {\bibfnamefont {C.~R.}\
  \bibnamefont {Dean}}, \ and\ \bibinfo {author} {\bibfnamefont {A.~F.}\
  \bibnamefont {Young}},\ }\href@noop {} {\bibfield  {journal} {\bibinfo
  {journal} {arXiv:1902.00763}\ } (\bibinfo {year} {2019})}\BibitemShut
  {NoStop}%
\bibitem [{\citenamefont {Cao}\ \emph {et~al.}(2016)\citenamefont {Cao},
  \citenamefont {Luo}, \citenamefont {Fatemi}, \citenamefont {Fang},
  \citenamefont {Sanchez-Yamagishi}, \citenamefont {Watanabe}, \citenamefont
  {Taniguchi}, \citenamefont {Kaxiras},\ and\ \citenamefont
  {Jarillo-Herrero}}]{Cao2016}%
  \BibitemOpen
  \bibfield  {author} {\bibinfo {author} {\bibfnamefont {Y.}~\bibnamefont
  {Cao}}, \bibinfo {author} {\bibfnamefont {J.~Y.}\ \bibnamefont {Luo}},
  \bibinfo {author} {\bibfnamefont {V.}~\bibnamefont {Fatemi}}, \bibinfo
  {author} {\bibfnamefont {S.}~\bibnamefont {Fang}}, \bibinfo {author}
  {\bibfnamefont {J.~D.}\ \bibnamefont {Sanchez-Yamagishi}}, \bibinfo {author}
  {\bibfnamefont {K.}~\bibnamefont {Watanabe}}, \bibinfo {author}
  {\bibfnamefont {T.}~\bibnamefont {Taniguchi}}, \bibinfo {author}
  {\bibfnamefont {E.}~\bibnamefont {Kaxiras}}, \ and\ \bibinfo {author}
  {\bibfnamefont {P.}~\bibnamefont {Jarillo-Herrero}},\ }\href@noop {}
  {\bibfield  {journal} {\bibinfo  {journal} {Physical Review Letters}\
  }\textbf {\bibinfo {volume} {117}} (\bibinfo {year} {2016})}\BibitemShut
  {NoStop}%
\bibitem [{\citenamefont {Bistritzer}\ and\ \citenamefont
  {MacDonald}(2011)}]{Bistritzer2011}%
  \BibitemOpen
  \bibfield  {author} {\bibinfo {author} {\bibfnamefont {R.}~\bibnamefont
  {Bistritzer}}\ and\ \bibinfo {author} {\bibfnamefont {A.~H.}\ \bibnamefont
  {MacDonald}},\ }\href@noop {} {\bibfield  {journal} {\bibinfo  {journal}
  {Proceedings of the National Academy of Sciences}\ }\textbf {\bibinfo
  {volume} {108}},\ \bibinfo {pages} {12233} (\bibinfo {year}
  {2011})}\BibitemShut {NoStop}%
\bibitem [{\citenamefont {Yuan}\ and\ \citenamefont {Fu}(2018)}]{Yuan2018}%
  \BibitemOpen
  \bibfield  {author} {\bibinfo {author} {\bibfnamefont {N.~F.~Q.}\
  \bibnamefont {Yuan}}\ and\ \bibinfo {author} {\bibfnamefont {L.}~\bibnamefont
  {Fu}},\ }\href@noop {} {\bibfield  {journal} {\bibinfo  {journal} {Physical
  Review B}\ }\textbf {\bibinfo {volume} {98}},\ \bibinfo {pages} {045103}
  (\bibinfo {year} {2018})}\BibitemShut {NoStop}%
\bibitem [{\citenamefont {Lian}\ \emph {et~al.}(2018)\citenamefont {Lian},
  \citenamefont {Xie},\ and\ \citenamefont {Bernevig}}]{Lian2018}%
  \BibitemOpen
  \bibfield  {author} {\bibinfo {author} {\bibfnamefont {B.}~\bibnamefont
  {Lian}}, \bibinfo {author} {\bibfnamefont {F.}~\bibnamefont {Xie}}, \ and\
  \bibinfo {author} {\bibfnamefont {B.~A.}\ \bibnamefont {Bernevig}},\
  }\href@noop {} {\bibfield  {journal} {\bibinfo  {journal} {arXiv:1811.11786}\
  } (\bibinfo {year} {2018})}\BibitemShut {NoStop}%
\bibitem [{\citenamefont {Ho}\ and\ \citenamefont {Li}(2019)}]{HL1}%
  \BibitemOpen
  \bibfield  {author} {\bibinfo {author} {\bibfnamefont {T.-L.}\ \bibnamefont
  {Ho}}\ and\ \bibinfo {author} {\bibfnamefont {C.}~\bibnamefont {Li}},\
  }\href@noop {} {\bibfield  {journal} {\bibinfo  {journal} {to be published}\
  } (\bibinfo {year} {2019})}\BibitemShut {NoStop}%
\end{thebibliography}%

\end{document}